\documentclass[preprint]{revtex4-1}
\usepackage{graphicx}
\usepackage{color}

\renewcommand\figurename{Figure}

\begin{document}

\title{Universality of fractal to non-fractal morphological transitions\\ in stochastic growth processes}
\author{J. R. Nicol\'as-Carlock}
\affiliation{Instituto de F\'isica, Benem\'erita Universidad Aut\'onoma de Puebla, Apdo.\ Postal.\ J-48, Puebla, Pue.\ 72570, Mexico.}
\author{J. L. Carrillo-Estrada}
\affiliation{Instituto de F\'isica, Benem\'erita Universidad Aut\'onoma de Puebla, Apdo.\ Postal.\ J-48, Puebla, Pue.\ 72570, Mexico.}
\author{V. Dossetti}
\affiliation{CIDS-Instituto de Ciencias, Benem\'erita Universidad Aut\'onoma de Puebla, Av. San Claudio esq. 14 Sur, Edif. 103D, Puebla, Pue.\ 72570, Mexico.}

\begin{abstract}

Stochastic growth processes give rise to diverse intricate structures everywhere and across all scales in nature. Despite the seemingly unrelated complex phenomena at their origin, the Laplacian growth theory has succeeded in unifying their treatment under one framework, nonetheless, important aspects regarding fractal to non-fractal morphological transitions, coming from the competition between screening and anisotropy-driven forces, still lacks a comprehensive description. Here we provide such unified description, encompassing all the known characteristics for these transitions, as well as new universal ones, through the statistical mix of basic models of particle-aggregation and the introduction of a phenomenological physically meaningful dimensionality function, that characterizes the fractality of a symmetry-breaking process induced by a generalized anisotropy-driven force. We also show that the generalized Laplacian growth (dielectric breakdown) model belongs to this class. Moreover, our results provide important insights on the dynamical origins of mono/multi-fractality in pattern formation, that generally occur in far-from-equilibrium processes.

\end{abstract}

\maketitle

\section*{Introduction}

From the formation of snowflakes to lightning, from mineral veins to bacterial colonies, the theory of Laplacian growth with its generalizations and extensions has highly contributed to our current understanding of far-from-equilibrium growth phenomena \cite{vicsek1992book, meakin1998book, vicsek2001book}. In particular, one striking phenomenological feature of this kind of systems is the morphological changes they undergo as a result of the interplay of the entropic, energetic and symmetry elements in their growth dynamics. This trends have been widely observed experimentally and, in some cases, successfully reproduced by computer simulations so that seemingly unrelated patterns found in nature are now understood in terms of a single generalized framework of complex growth \cite{benjacob1990, benjacob1993, benjacob1997}. However, complexity in nature seems to follow non-trivial paths revealed in self-organizing and self-assembling processes that in most cases are characterized by critical and/or morphological transitions that cannot be properly described by the Laplacian growth models \cite{sander1986, sander2000, sander2011}. 

In the Laplacian theory, the probability of growth at a given point in space, $\mu$, is proportional to the spatial variation of a scalar field, $\phi$, with $\mu\propto |\nabla\phi|$. One example of such processes is the paradigmatic diffusion-limited aggregation (DLA) model, where particles following Brownian trajectories aggregate one-by-one to form a cluster, starting with a seed particle \cite{sander2000, sander2011} (see Fig.\ 1). It has been found that the disordered structure that emerges from this kind of processes shows self-similarity, described by a single fractal dimension $D$ \cite{sander2011}. Mean-field analyses have shown that this fractal dimension depends only on the dimension $d$ of the space where the cluster grows as, $D(d)=(d^2+1)/(d+1)$ \cite{muthukumar1983, tokuyama1984}. In two dimensions, this expression predicts $D=5/3\approx 1.67$, that is not very far from the widely reported value for off-lattice DLA clusters, $D=1.71$. Furthermore, the fact that the fractal features of these clusters are highly dependent on the mean square displacement of the particles in the trajectories they follow before aggregation, gives rise to a continuous screening-driven morphological transition that has been neatly described by extending the Laplacian theory to consider a general aggregation process where particles follow fractal trajectories \cite{meakin1984a}. The cluster's fractal dimension is then related to that of the walkers' trajectories, $d_w$, through the Honda-Toyoki-Matsushita (HTM) mean-field equation, $D(d,d_w)=(d^2+d_w-1)/(d+d_w-1)$, where $d$ is the dimension of the embedding space \cite{matsushita1986a, matsushita1986b}. Here, for $d_w=1$ one gets $D=d$, as expected for ballistic-aggregation (BA) dynamics, whereas for $d_w=2$, the value $D=5/3$ for DLA is recovered. This kinetically induced BA-DLA transition has been reproduced in diverse and equivalent aggregation schemes, for example, under drift of wandering particles \cite{meakin1983}, using particles with variable random-walk step size \cite{huang1987}, by imposing directional correlations \cite{huang2001, ferreira2005}, and through probabilistically mixed-dynamics aggregation \cite{alves2006}. 

Nonetheless, one of the most challenging aspects of the theory arises when the growth is not purely limited by diffusive processes, for example, when it takes place under the presence of long-range attractive interactions or under the effects of surface tension, where strong screening and anisotropic effects must be taken into account \cite{benjacob1990, benjacob1993}. For this, a clever generalization to the diffusion-limited growth processes was proposed within the context of the dielectric-breakdown (DB) model, assuming $\mu\propto |\nabla(\phi)|^\eta$, where $\eta$ is a positive real number associated with non-linear effects coming from screening and anisotropy \cite{pietronero1984}, while a very good description of the self-similarity in the emerging structures was provided by the generalized HTM equation (GHTM) \cite{matsushita1986b, matsushita1986c},
\begin{equation}
D(d,d_w,\eta)=\frac{d^2+\eta(d_w-1)}{d+\eta(d_w-1)}.
\end{equation} 
This equation predicts a continuous morphological transition, from a compact structure with $D=d$ (BA) as $\eta \to 0$, to a one-dimensional one in the highly anisotropic regime, i.e., $D=1$ as $\eta\to\infty$. In particular, for two-dimensional diffusive-type processes ($d=d_w=2$), the value $D=5/3$ is recovered for $\eta=1$. Due to the limitations of this expression, coming from its mean-field approach, extensive numerical work has been invested to prove the transition from branching fractals to one-dimensional non-fractal clusters at the critical value $\eta_c\geq 4$ \cite{sanchez1993, hastings2001, mathiesen2002}. The use of the fractal dimension as an order parameter to describe these transitions is still debatable as well. These are some of the important aspects we will address in this Article.

\begin{figure}[tb]
\includegraphics[width=0.55\textwidth]{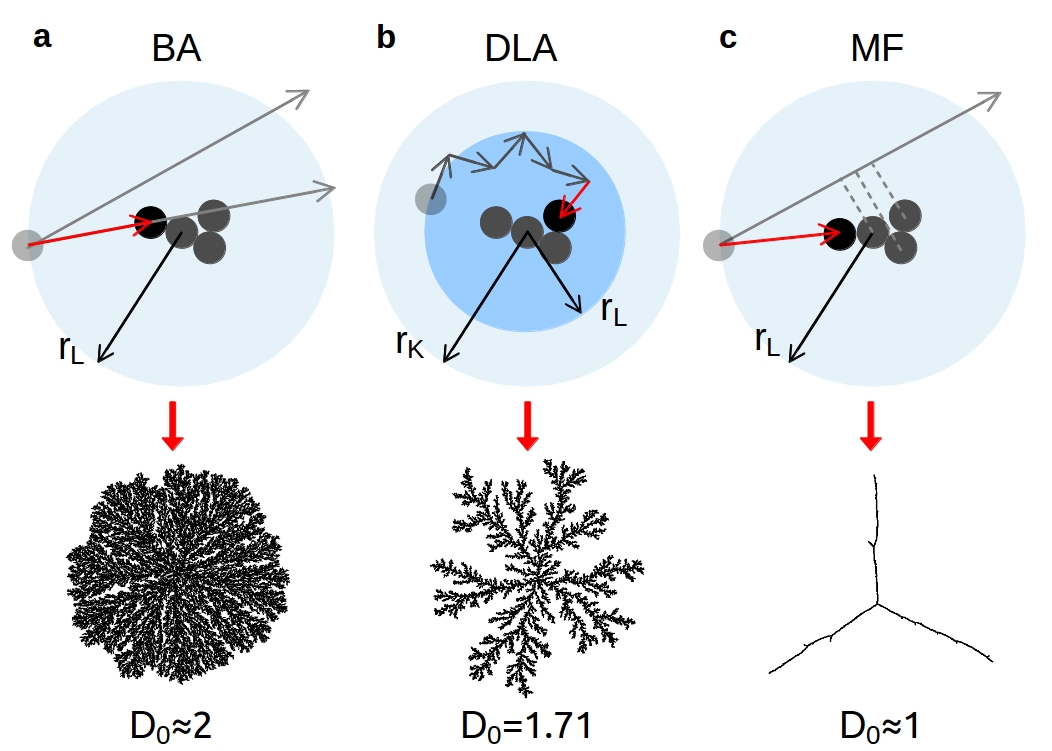}
\caption{\label{fig1} \textbf{Fundamental models of aggregation.} After been launched into the system from $r_L$ with uniform probability, particles (\textbf{a}) follow straight-line trajectories before aggregation takes place in BA, (\textbf{b}) perform a random walk in DLA, and (\textbf{c}) get radially attached to the closest particle in the cluster as a result of an infinite-range (system-size) radial interaction in MF aggregation. The latter is particle-path independent and its morphological characteristics emerge solely from this long-range interaction as opposed to the stochastic BA and DLA models. The characteristic fractal dimension $D_0$ for each type of aggregation process is indicated.}
\end{figure}

\begin{figure*}[htb]
\includegraphics[width=\textwidth]{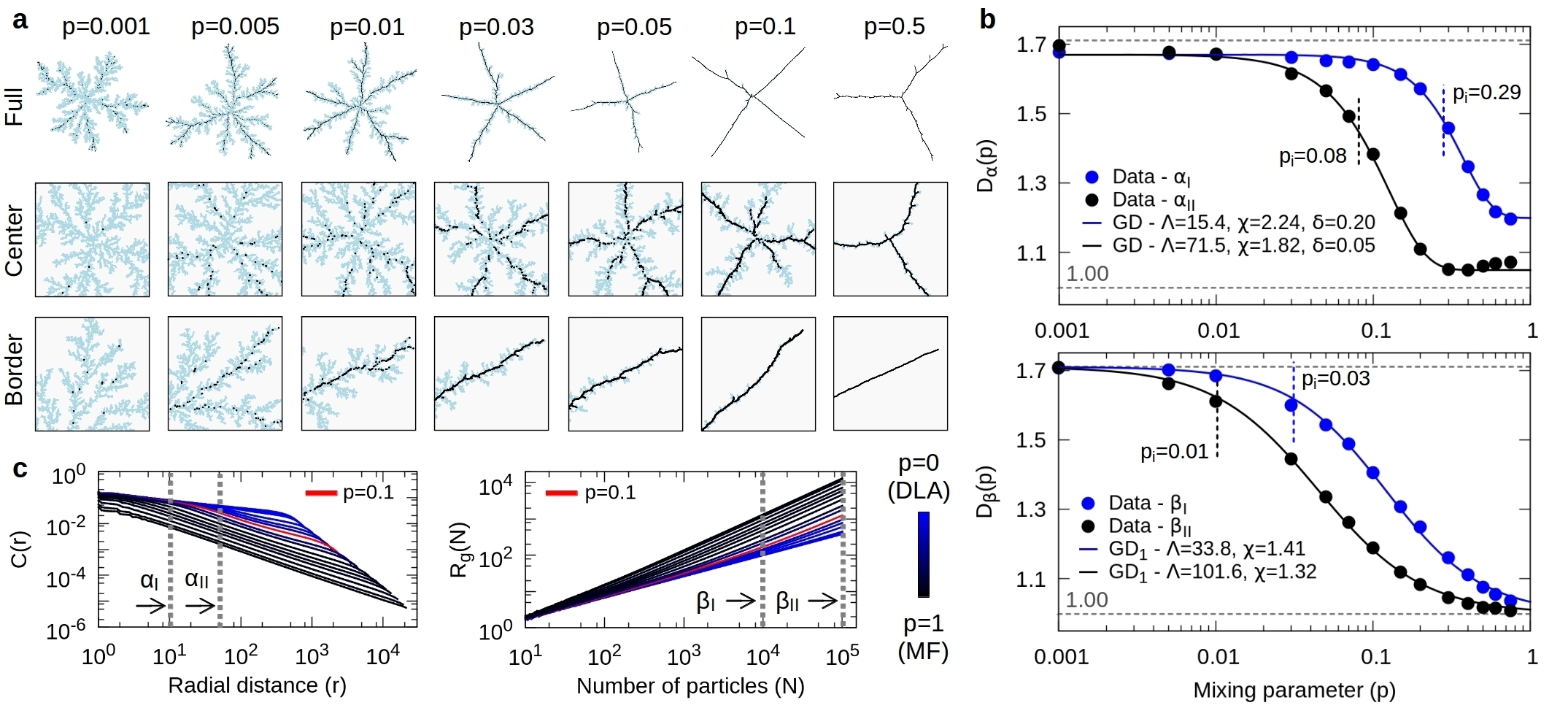}
\caption{\label{fig2} \textbf{DLA-MF transition.} As shown in (\textbf{a}), this transition is characterized by a remarkably fast morphological evolution from unstable tip-splitting, through dendritic growth, to needle-like (MF) growth as $p$ increases. Particles aggregated under DLA dynamics are coloured in light-blue while those through MF dynamics in black. In (\textbf{b}), the measured $D_{\alpha}(p)$ and $D_{\beta}(p)$ are fitted using the GD and GD$_1$ functions with $\Lambda$ and $\chi$ as the fitting parameters, while the computed $p_\mathrm{i}$ for each curve is marked with vertical dotted lines. In (\textbf{c}), $C(r)$ and $R_g(N)$ display deviations from a well-defined linearity over their respective scales and for different values of $p$, revealing the inhomogeneity of these clusters.}
\end{figure*}

\section*{The model}

For our numerical analysis we considered a combinatorial (Monte Carlo) scheme of three fundamental and simple two-dimensional \textsl{off-lattice} models of particle-cluster aggregation. On the one hand, the DLA and BA models will provide us with disordered/fractal structures through the full range of stochastic aggregation dynamics (Figs.\ 1a and 1b). On the other, the long-range particle-cluster interactive mean-field (MF) model will be an agent of order, providing the most energetic (or noiseless) aggregation dynamics \cite{jullien1986, nicolas2016}, as well as acting as the main source of anisotropy, in this case purely generated by the growth dynamics but not from lattice effects \cite{meakin1988} (see Fig.\ 1c). The fractional combination of these models, controlled by the mixing parameter $p\in[0,1]$, results in DLA-MF and BA-MF transitions, going from homogeneous fractal structures with $d \geq D > 1$ when $p \to 0$, passing through inhomogeneous multifractal structures, to non-fractal structures with $D=1$ as $p \to 1$. Before discussing these results in details in the next section, let us develop a general framework to analyze morphological transitions in stochastic growth processes.

Using the fractal dimension of the clusters $D$, to characterize a given morphological transition, let us define the function $f(p)=\Lambda p^\chi$ as a real and positive monotonically-increasing continuous function of $p$. Here, $f(p)$ takes a similar role as $\eta$ in the DB model, associated with the net effects of all screening/anisotropy-driven growth forces, that is, all the symmetry-breaking effects, where $\Lambda$ and $\chi$ are two po\-si\-ti\-ve real numbers associated with the strength of these forces, while $p$ modulates the anisotropy in such a way that $f(p)\to 0$ as $p\to 0$, and $f(p)\to\infty$ as $p\to 1$; $p\in[0,1]$ as before. These mathematical characteristics make plausible to define a general dimensionality (GD) function, $D(D_0,f(p))$, that describes the fractal Hausdorff dimension of a structure that collapses to $D=1$ under the effects of $f(p)$ as, 
\begin{equation}	
D(D_0,f(p))=1+(D_0-1)e^{-f(p)/D_0},
\end{equation}
where $D_0$ is the fractal dimension of clusters produced in the most isotropic and stochastic scenario, i.e., in the absence of any anisotropy-driven forces, while the exponential function allows us to consider all of the powers of $p$. Equation (2) is characterized by an inflection point, $p_\mathrm{i}$, that must satisfy $(\text{d}f/\text{d}p)^2=D_0(\text{d}^2f/\text{d}p^2)$ which, for the choice of $f(p)$, will be given by
\begin{equation}
p_\mathrm{i} = \biggl[\frac{D_0}{\Lambda}\biggl(\frac{\chi-1}{\chi}\biggr)\biggr]^{1/\chi}.
\end{equation} 

The first-order approximation (GD$_1$) of the general dimensionality function (2) can be written as,
\begin{equation}
D^{(1)}=\frac{D_0^2+f(p)}{D_0+f(p)},
\end{equation}
with an inflection point that must now satisfy $(\text{d}f/\text{d}p)^2=2(\text{d}^2f/\text{d}p^2)/(D_0+f)$, and is given by
\begin{equation}
p_\mathrm{i}^{(1)} = \biggl[\frac{D_0}{\Lambda}\biggl(\frac{\chi-1}{\chi+1}\biggr)\biggr]^{1/\chi}.
\end{equation} 

As shown, these expressions predict a continuous morphological transition from $D \to D_0$ as $f(p)\to 0$ (disordered/fractal state) towards $D \to 1$ as $f(p) \to \infty$ (ordered state), with an expected change in growth dynamics at $p_\mathrm{i}$. Additionally, we will also define the reduced dimensionality (RD) function as the transformation $D^*=(D-1)/(D_0-1)$, with $D^*\in[0,1]$, that leaves the inflection-point of equations (3) and (5) invariant, given by
\begin{equation}
D^*=e^{-f(p)/D_0},
\end{equation}
the first-order approximation (RD$_1$) becomes,
\begin{equation}
D^{*(1)}=\frac{1}{1+f(p)/D_0},
\end{equation}
where $f(p)$ is given as before.

In the following section we will use these expressions to analyze some important cases.

\section*{Results and discussion}

For all of our numerically generated clusters, the fractal dimension was measured by means of two standard procedures: the two-point density radial correlation function $C(r)$ and the radius of gyration $R_g(N)$, that yield $D_\alpha$ and $D_\beta$, respectively (see Methods for more details).

\begin{figure*}[htb]
\includegraphics[width=\textwidth]{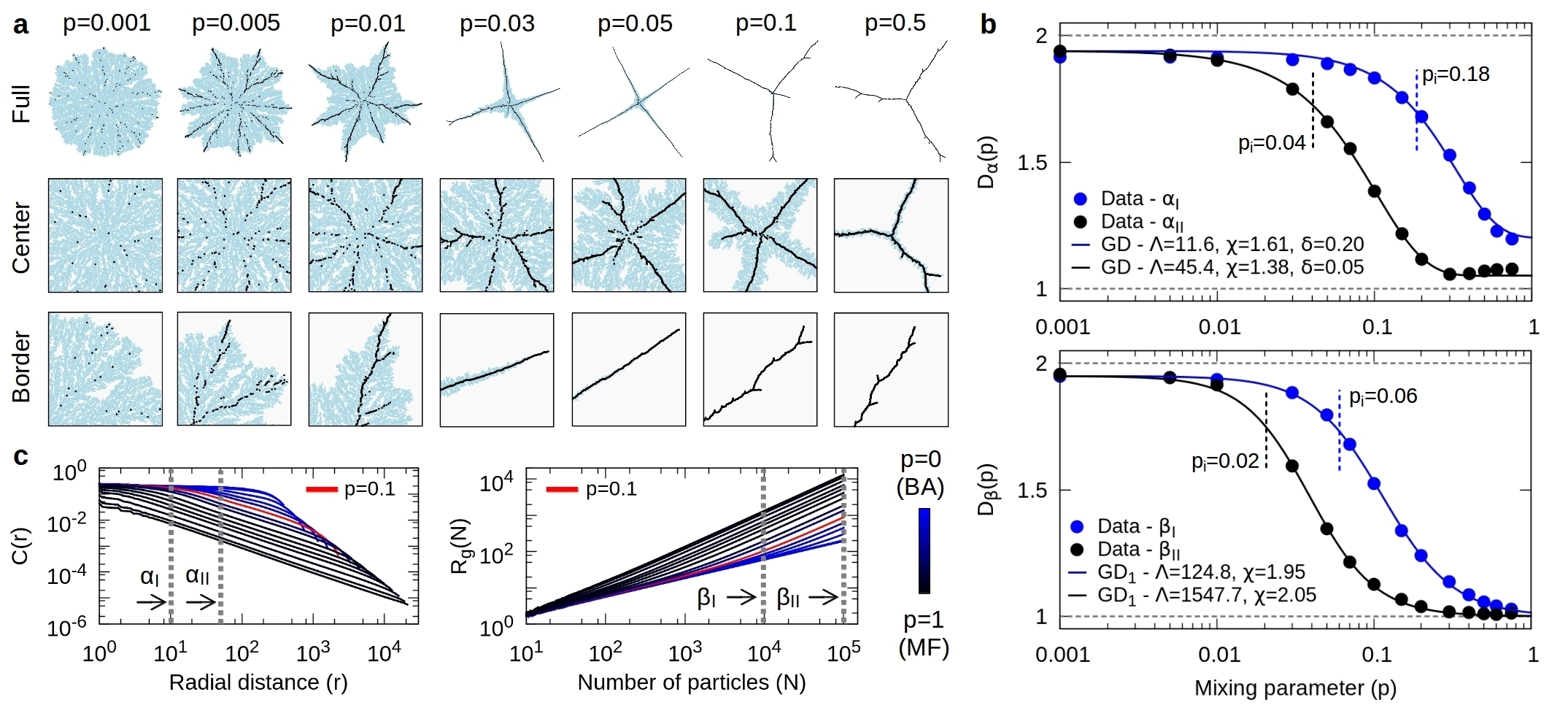}
\caption{\label{fig3} \textbf{BA-MF transition.} As shown in (\textbf{a}), this transition is characterized by a morphological evolution from dense branching, through dendritic growth, to needle-like (MF) growth, as $p$ increases. Particles aggregated under BA dynamics are coloured in light-blue while those through MF dynamics in black. In (\textbf{b}), the measured $D_{\alpha}(p)$ and $D_{\beta}(p)$ are fitted using the GD and GD$_1$ functions with $\Lambda$ and $\chi$ as the fitting parameters, while the computed $p_\mathrm{i}$ for each curve is marked with vertical dotted lines. In (\textbf{c}), $C(r)$ and $R_g(N)$ display deviations from a well-defined linearity over their respective scales and for different values of $p$, also revealing the inhomogeneity of these clusters.}
\end{figure*}

\begin{figure*}[htb]
\includegraphics[width=\textwidth]{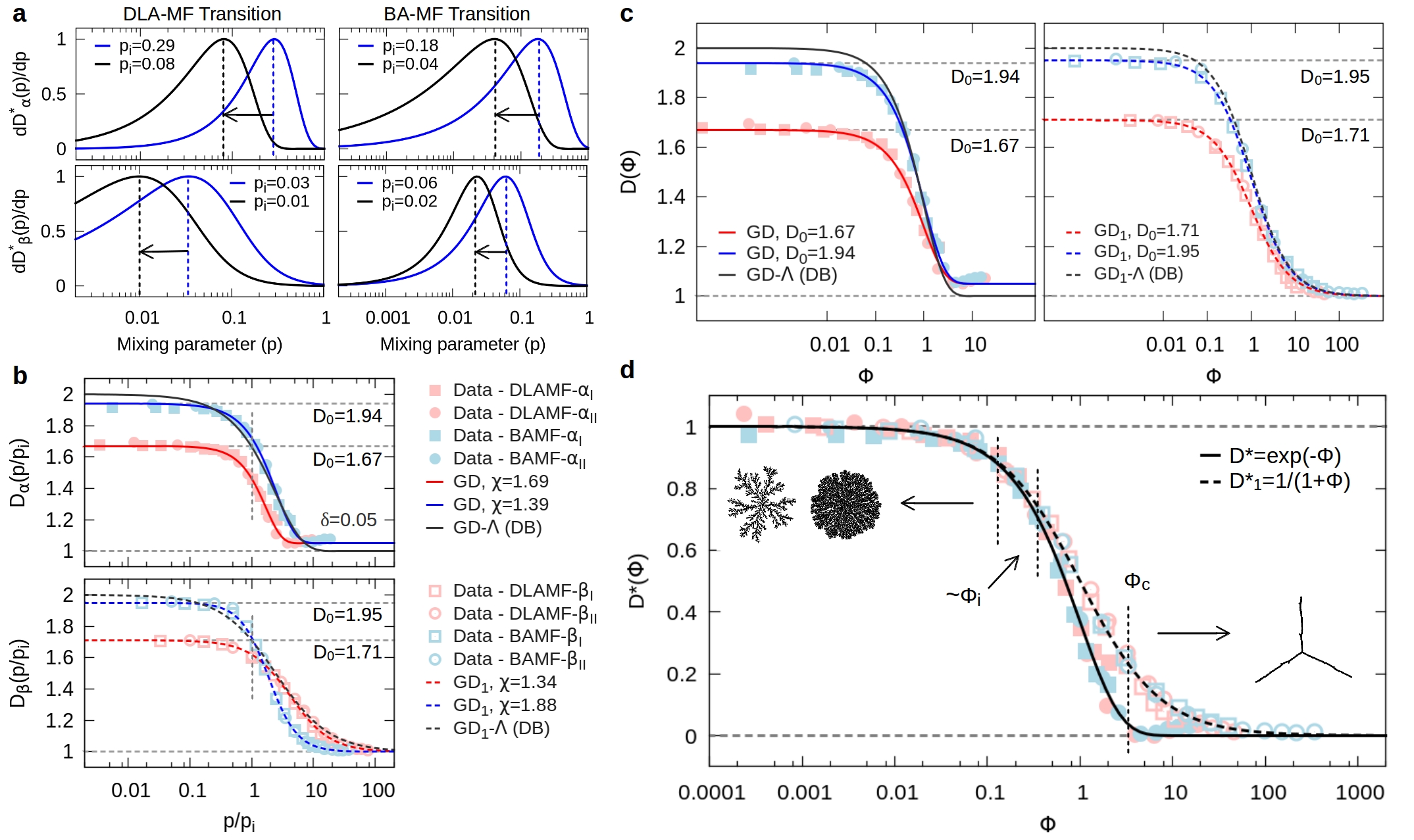}
\caption{\label{fig5} \textbf{Universality.} (\textbf{a}) First derivatives of $D^*_\alpha(p)$ and $D^*_\beta(p)$ for DLA-MF and BA-MF transitions, normalized to their maximum value located at their respective $p_\mathrm{i}$. (\textbf{b}) By plotting $D_\alpha$ and $D_\beta$ as a function of $q=p/p_i$, all of the data collapse into single curves. The curves GD-$\Lambda$ and GD$_1$-$\Lambda$ for the DB model are also included. (\textbf{c}) Plotting $D_\alpha$ and $D_\beta$ as function of $\Phi$ shows that all transitions share a common critical point $\Phi_c$ . (\textbf{d}) Plotting $D^*_\alpha$ and $D^*_\beta$ data as function of $\Phi$, the DLA-MF and BA-MF transitions become independent of the fractal dimension of the  stochastic model considered, described by the same (solid) curve. Under this transformation, GD-$\Lambda$ and GD$_1$-$\Lambda$ also become $D^*$ and $D^*_1$, respectively (see text for more details).}
\end{figure*}

\subsection*{DLA-MF and BA-MF transitions}

In the anisotropy-driven DLA-MF and BA-MF transitions, the variable $p$ is associated with the fraction of particles aggregated under MF dynamics, that is $p=N_\mathrm{MF}/N$, where $N$ is total number of particles. Therefore, as $p$ varies from $p=0$ (DLA or BA) to $p=1$ (MF), it controls the morphology of the system through a continuous and extremely fast symmetry-breaking induced by the anisotropy of the MF dynamics (see Figs.\ 2a and 3a). Contrary to the screening-driven BA-DLA transition \cite{meakin1984a, ferreira2005}, the clusters generated by these processes are inhomogeneous fractals, this is, structures with different fractal dimensionality at different scales (see Figs.\ 2b, 2c, 3b, and 3c). Even more, these transitions are characterized by fast and well-defined changes in growth dynamics as $p$ changes, therefore, they can no longer be properly described by the GHTM model. In contrast, the proposed GD function can be used to estimate the measured fractal dimensions of the clusters at different scales, using $\Lambda$ and $\chi$ as fitting parameters. In fact, we found that numerically obtained data for $D_\alpha(p)$ are well described by the GD function, whereas GD$_1$ describes best the dependence of $D_\beta(p)$ (Fig.\ 2b and 3b). Furthermore, even though it is possible to define the inflection points for each curve, the inhomogeneity of the clusters makes it impossible to establish a well-defined morphological critical point in the same manner as for the DB model, that is, a point where highly anisotropic effects are dominant in determining the morphology of the clusters \cite{sanchez1993, hastings2001}. Nonetheless, such a description is possible in our new framework under proper rescaling as explained later.

One important implication that can be drawn from the previous analysis, comes from the role of $\Lambda$ in des\-cri\-bing the fractality of these clusters. In the BA-DLA transition, both $\Lambda$ and $\chi$ are constant, leading to homogeneous fractals \cite{meakin1983, huang1987, huang2001,ferreira2005, alves2006}. This is seen by rewriting the GHTM equation in terms of $p$ as, $D(d,p,\eta)=(d^2+\eta p)/(d+\eta p)$, with $p=d_w-1$. The HTM equation that describes the BA-DLA transition, is recovered for $\Lambda = \chi = \eta =1$. This is also the case for the DB model where $\Lambda = \chi = 1$ \cite{pietronero1984, matsushita1986c, sanchez1993}. In contrast, these parameters have different values in the DLA-MF and BA-MF transitions, depending on the scale in which $D_\alpha(p)$ and $D_\beta(p)$ are measured (see Figs.\ 2b and 3b). This suggest that, in order to have a transition characterized by homogeneous fractals, the amplitude of the screening/anisotropic force, $\Lambda$, must remain constant along the transition and across different scales. This is also supported by noticing that $\Lambda$ will be bounded by the limit value of the GD function when $p\to 1$, i.e., when all of the particles follow MF dynamics. In this limit, $D_{p\to 1}=1+(D_0-1)\exp(-\Lambda/D_0)$, must satisfy $D_{p\to 1}=1$ (from its definition), which would lead to $\Lambda/D_0\gg 1$. This is a convergence condition in the highly anisotropic regime that is better defined by considering $D_{p\to 1}\leq 1+\delta$, where $\delta\ll 1$, is a measure of the deviation from a given structure to fully collapsing into a one-dimensional one. This condition leads to $\Lambda\geq -D_0\log[\delta/(D_0-1)]$, that establishes a lower bound, not only for $\Lambda$, but for $D(p)$ as well. Thus, given a $D_0$ (also bounded by $d$, i.e., the dimension of the embedding space), $\Lambda$ has a lower bound that depends only on $\delta$. This implies that long-correlated structures are not restricted to develop a single scaling-law or mono-fractal features, and that inhomogeneous or \emph{multi-fractal structures} are more likely to arise in these out-of-equilibrium growth processes. Further evidence supporting this observation is provided bellow.

\subsection*{Universality}

The proposed GD function and its first-order approximation are powerful phenomenological expression, able to describe the fractality of the clusters at different scales and along the transitions, as well as the changes in growth regimes through the inflection points (see Fig.\ 4a). However, a full insight into the nature of these transitions is needed, in particular, the possibility of defining morphological critical points (as previously done for the DB model in \cite{sanchez1993, hastings2001}). To this end, let us notice that under the scaled variable, $q=p/p_i$, it is possible to collapse all the data for $D(p)$ into a single curve according to their respective transition and measurement method (see Fig.\ 4b). This implies that the symmetry-breaking induced by the anisotropy-driven force is a unique process whose manifestations are the same across all of the scales. Moreover, given the uniqueness of these processes, we are in the possibility of defining critical points by substituting $q\in[0,\infty)$, back into the GD function, leading to,
\begin{equation}	
D(D_0,\Phi(q))=1+(D_0-1)e^{-\Phi(q)},
\end{equation}
where $\Phi(q)=\Lambda(\chi)q^\chi$, with $\Lambda(\chi)=(\chi-1)/\chi$, is associated to a generalized anisotropy-driven force. Its first-order approximation (GD$_1$) is given by, 
\begin{equation}	
D(D_0,\Phi(q))^{(1)}=\frac{D_0+\Phi(q)}{1+\Phi(q)},
\end{equation}
where $\Phi(q)=\Lambda(\chi)q^\chi$, with $\Lambda(\chi)=(\chi-1)/(\chi+1)$. By construction, all inflection points are now located at $q_i=1$. Fitting these functions to $D_\alpha(q)$ and $D_\beta(q)$, respectively, we obtain the curves shown in Fig.\ 4b, that are in excellent agreement with the data. In particular, in this approach $\Lambda$ depends only on $\chi$, thus providing further evidence to support our argument about the origins of mono or multi-fractality in these morphological transitions.

Furthermore, by plotting all data for $D_\alpha(q)$ and $D_\beta(q)$ as function of $\Phi(q)$, it can be observed that both the DLA-MF and BA-MF transitions equally approach the highly anisotropic regime, showing that the symmetry-breaking process driven by $\Phi$ is independent of the initial configuration of the system, as shown in Fig.\ 4c. This can be better appreciated by using the reduced dimensionality transformation to plot $D^*_\alpha$ and $D^*_\beta$ as function of $\Phi$ (see Fig.\ 4d), under which one can observe the full collapsing of the data into single universal curves that are perfectly described by the reduced dimensionality (RD) function and its first-order approximation (RD$_1$) given respectively by, 
\begin{equation}
D(\Phi)^*=e^{-\Phi(q)}, \quad D(\Phi)^{*(1)}=\frac{1}{1+\Phi(q)}.
\end{equation}	

In fact, since $D_0$ can be well described by the HTM model as $D_0(d,d_w)=(d^2+d_w-1)/(d+d_w-1)$, and given that these symmetry-breaking processes are initial-configuration independent, we might conclude that these fractal to non-fractal morphological transitions will depend solely on the strength of the anisotropic-driven force acting upon them, following the same universal fractality curves in any dimension.

Regarding this universality concept, we must recall that the dielectric breakdown and even viscous fingering phenomena are said to belong to the same universality class as DLA, because they are characterized by the same fractal dimension, generated by related Laplacian growth processes, and not because any other thermodynamic criteria \cite{sander2011, mathiesen2006}. Therefore, the \textsl{universality} of these morphological DLA-MF and BA-MF transitions must be understood in the sense as they are described by the same general equations in the $D^*(\Phi)$-space, independent of the dimensionality of their embedding space. Note that due to this universality feature, in this reduced space these transitions can be seen as a regular phase-transition. For this, let us define the \textsl{critical point} of these transitions analogously to that of the DB model, i.e., as the point where the fractal dimension of the clusters is slightly different from the dimension of a one-dimensional structure \cite{hastings2001}. This can be achieved by considering $D^*(\Phi_c)=\epsilon$, where $\epsilon\ll 1$. Therefore, from the RD and RD$_1$ functions, the \textsl{universal critical points}, $\Phi_c$, must satisfy, $\exp(-\Phi_c)=\epsilon$ and $\Phi_c=(1-\epsilon)/\epsilon$, respectively. In order to recover the particular critical points for each transition, we must recall that $\Phi_c=\Phi(q_c)$ then, one has to solve for $q_c$ accordingly. Notice also that in this case, $q_c$ depends on $\epsilon$, $\chi$ and $D_0$, therefore, giving different values for each transition (Table I).

\subsubsection*{Universality of the DB model}

As previously stated, the GHTM mean-field equation that describes the fractality of clusters generated in the DB model gives $D=5/3\approx 1.67$, for $\eta=1$ (DLA). This can be corrected by noticing that equation (1), re-written as $D(d,\eta)=(d+\eta/d)/(1+\eta/d)$, is similar to the description given by the first-order approximation equation $D(D_0,\Phi)^{(1)}$, with the peculiarity that, in this case, $\Phi\sim\eta/d$. Considering $\Phi(\eta)=\Lambda\eta/d$, the corrected GHTM equation (GD$_1$-$\Lambda$) is given by, $D(d,\eta)^{(1)}=(d+\Lambda\eta/d)/(1+\Lambda\eta/d)$. In two dimensions, setting $\eta=1$ and $D=1.71$ yields $\Lambda=0.817$. Furthermore, this equation can now be identified as the first-order approximation of a general function (GD-$\Lambda$) given by, $D(d,\eta)=1+(d-1)\exp(-\Lambda\eta/d)$ which, for $d=2$, is in excellent agreement with previous numerical results \cite{matsushita1986c, pietronero1988, hastings2001} (see Table II). This DB transition draws interesting similarities with the BA-MF transition, since both of them exhibit a symmetry-breaking from $D=d$ to $D=1$, have critical points close to $4$, and $D\approx 1.71$, for $q=1$ and $\eta=1$, even though we are dealing with completely different growth processes (see Fig.\ 4b and Table I).

On the other hand, since the DB model is associated to a generalized Laplacian process described by $\mu\propto|\nabla(\phi)|^{\eta}$, one would be tempted to associate the BA-MF model to a similar processes where $\eta \to q$. However, it will not be correct because both processes would become equivalent and the numerical solution of the equation would result in the generation of equivalent morphological clusters, which is obviously incorrect. Moreover, another important difference between the DB and BA-MF description, is that the case $\eta=1$ will not be associated with any inflection point in the dynamics, as it is the case for $q_i=1$, since the condition $\chi>1$ is not satisfied, as can be seen from simple inspection of the GD function. Interestingly enough, if the case $\eta=1$ could have been directly related to the case $q_i=1$, this would have indicated that the DLA process is directly associated to an inflection point in the dynamics and not to a critical point as was suggested through a completely different approach \cite{dimino1989, kaufman1989}. Worthy of remark is that our approach provides the corrected form for all the previous mean-field equation for the fractality of the generalized Laplacian processes, within the first-order approximation of the general GD-$\Lambda$ equation, given by,
\begin{equation}
D(d,d_w,\eta)=1+(d-1)e^{-\Lambda\eta(d_w-1)/d},
\end{equation} 
whose first-order approximation, GD$_1$-$\Lambda$, is given by, 
\begin{equation}
D(d,d_w,\eta)^{(1)}=\frac{d+\Lambda\eta(d_w-1)/d}{1+\Lambda\eta(d_w-1)/d},
\end{equation}
where, $\Lambda=0.817\approx 4/5$, thus unifying all of the models. 
As such, the corrected GHTM equation for the DB model is recovered by setting $d_w=2$; the corrected HTM equation for the BA-DLA transition is recovered for $\eta=1$; and the corrected DLA mean-field equation is recovered for $d_w=2$ and $\eta=1$. Finally and importantly, under the reduced dimensionality transformation, equations GD-$\Lambda$ and GD$_1$-$\Lambda$ are respectively given by $D(\Phi(\eta))^*=\exp(-\Phi)$ and $D(\Phi(\eta))^{*(1)}=1/(1+\Phi)$, with $\Phi=\Lambda\eta(d_w-1)/d$, showing that they belong to the same universality class as the DLA-MF and BA-MF transitions (see Figs.\ 4c and 4d).

\begin{table*}[htb]
\caption{\label{critical}\textbf{Critical points.} Estimated $\Phi_c$ and $q_c$, using the respective values of $D_0$, $\chi$ and $\epsilon$. Attention should be paid to the BAMF-$\alpha$ transition, that exhibits remarkable similarities with the DB transition \cite{sanchez1993,hastings2001}, such as, a $q_c\approx 4$ and $D\approx 1.71$ at $q=1$, which would be similar to the case $\eta=1$, even though these transitions come from different processes.}
\begin{ruledtabular}
\begin{tabular}{lccccccc}
Data&$D_0$&$\chi$&$\Phi_c(\epsilon=0.1$)&$q_c$&$\Phi_c(\epsilon=0.05)$&$q_c$&$D(q=1)$\\
\hline
BAMF-$\alpha$	&$1.9384\pm 0.0001$	& $1.39\pm 0.02$ 	& $2.3$	& $4.5$	& $3.0$ & $5.4$ & $1.72\pm 0.02$\\
DLAMF-$\alpha$	&$1.6749\pm 0.0024$	& $1.69\pm 0.02$ 	& $2.3$	& $2.8$ & $3.0$ & $3.2$ & $1.46\pm 0.02$\\
BAMF-$\beta$	&$1.9485\pm 0.0001$	& $1.88\pm 0.01$ 	& $9.0$	& $6.0$ & $19.0$ & $9.0$ & $1.73\pm 0.01$\\
DLAMF-$\beta$	&$1.7100\pm 0.0007$	& $1.34\pm 0.01$ 	& $9.0$	& $21.7$ & $19.0$ & $37.8$ & $1.62\pm 0.01$\\
\end{tabular}
\end{ruledtabular}
\end{table*}

\begin{table}[htb]
\caption{\label{critical2}\textbf{DB equations comparison.} Fractal dimensions obtained for the DB model ($d=2$) at different values of $\eta$, as reported in \cite{hastings2001}, with the values estimated by the GHTM equation, and the corrected GD-$\Lambda$ and GD$_1$-$\Lambda$ equations, with $\Lambda=0.817$. In particular, GD-$\Lambda$ is in great agreement with the reported values for DB within the numerical uncertainties.}
\begin{ruledtabular}
\begin{tabular}{lcccc}
$\eta$	&	DB	& GHTM	& GD$_1$-$\Lambda$	&	GD-$\Lambda$ \\
\hline
1	&-					& $5/3\approx 1.667$	& $1.710$	& $1.665$\\
2	&$1.433\pm 0.039$	& $3/2=1.500$			& $1.550$	& $1.442$\\
3	&$1.263\pm 0.056$	& $7/5=1.400$			& $1.449$	& $1.294$\\
4	&$1.128\pm 0.072$	& $4/3\approx 1.333$	& $1.379$	& $1.195$\\
5	&$1.068\pm 0.046$	& $9/7\approx 1.286$	& $1.328$	& $1.129$\\
\end{tabular}
\end{ruledtabular}
\end{table}

On a separate note, given that in all of these transitions $D_\alpha$ and $D_\beta$ are finelly well described by the RD and RD$_1$ functions, respectively, this indeed suggests that the true fractal dimension of DLA clusters is $D=1.67$, whereas $D=1.71$ is its value at first-order approximation. This observation is in good agreement with some results previously reported based on  robust methods such as the two-point density correlation function, while results based on fast methods such as the radius of gyration report values close to 1.71 \cite{meakin1984a, huang2001, alves2006, pietronero1984, meakin1983b, tolman1989, halsey1992, halsey1994, mandelbrot2002} (see also Extended Data).

\subsection*{Conclusions}

In this work we present a set of dimensionality functions that provide an useful and power\-ful description of the fractality in the anisotropy-driven DLA-MF and BA-MF transitions. Under a generalized anisotropy-force approach and a reduced dimensionality transformation, these transitions follow universal curves, showing that they are independent of their initial fractal dimension, and only dependent on the anisotropy force acting upon them. Also, provided that the initial fractal dimension is a function of the dimensionality of the embedding space (as described by the HTM equation), these results reveal the universality of these fractal to non-fractal morphological transitions. As well, we also show that the DB model (the generalized Laplacian model) belongs to the same universality-class, sharing similarities with the BA-MF model such as critical points and close values of their fractal dimension at $q_i=1$ or $\eta=1$. Additionally, we introduce a correction to the well-established GHTM mean-field approximation, that leads to a solution capable to quantitatively reproduce with remarkable precision previous well-established observations. In summary, we present here, for the first time, a comprehensive discussion on the dynamical origin of the fractality and, with this, the basis for the understanding its subjacent algebra originated in these far-from-equilibrium transitions as an emergent feature of the anisotropy effects. These results represent an important unifying step towards a complete theory of growth, and they might provide important insights to understand pattern formation phenomena in many related areas of research  \cite{vicsek2001book, benjacob1997, lehn2002, whitesides2002, sturmberg2013}.

\section*{Methods}

\subsection*{Aggregation dynamics}

In all simulations, each particle has a diameter equal to one. This is the basic unit of distance for the system. For aggregates based on BA or MF (Fig.\ 1a and 1c), we follow a standard procedure in which particles are launched at random from the circumference of radius $r_L = 2r_{max} + \delta$, with equal probability in position and direction of motion. Here, $r_{max}$ is the distance of the farthest particle in the cluster with respect to the seed particle placed at the origin. In our simulations we used $\delta=1000$ particle diameters to avoid undesirable screening effects. In particular, for the MF model, particles always aggregate to the closest particle in the cluster, this is determined by the projected position of the aggregated particles along the direction of motion of the incoming particle (see Fig.\ 1c). In the case of aggregates based on DLA (Fig.\ 1b), particles were launched from a circumference of radius $r_L = r_{max} + \lambda + \delta$, with $\delta=100$. The mean free path of the particles is set to one particle diameter. We also used a standard scheme that modifies the mean free path of the particles as they wander at a distance larger than $r_L$ or in-between branches, and set a killing radius at $r_K = 2r_L$, in order to speed up the aggregation process.

On the other hand, in order to mix different aggregation dynamics, an aggregation scheme is selected with probability $p$ while the other with probability $p-1$, as explained before. The evaluation of the aggregation scheme to be used is only updated once a particle has been successfully aggregated to the cluster under such dynamics.

\subsection*{Evaluating the fractal dimension}

In all measurements, we used $128$ clusters containing $10^5$ particles. The fractal dimension is measured from the two-point density correlation function, $C(r)=\langle\langle\rho(\textbf{r}_0)\rho(\textbf{r}_0+\textbf{r})\rangle\rangle_{|\textbf{r}|=r}$, where the double bracket indicates an average over all possible origins $\textbf{r}_0$ and all possible orientations. Here, it is assumed that $C(r)\approx r^{-\alpha}$, where the fractal dimension is given by $D_\alpha=d-\alpha$ where $d$ is the dimension of the embedding space. We also used the radius of gyration given by $R_g^ 2=\sum_{i=1}^{N}(\textbf{r}_i-\textbf{r}_{CM})^2$, where $N$ is the number of particles, $\textbf{r}_i$ is the position of the $ith$-particle in the cluster, and, $\textbf{r}_{CM}$ is the position of the center of mass. Here, it is assumed that $R_g(N)\approx N^\beta$, where the fractal dimension is given by $D_\beta=1/\beta$. Therefore, the fractal dimensions, $D_\alpha$ and $D_\beta$, are obtained from linear-fits to the corresponding functions, $C(r)$ and $R_g(N)$, in log-log plots. In practice, it is assumed that $\alpha$ and $\beta$ are constant as long as the size or number of particles in the cluster is large. However, because some clusters do not develop a constant fractal dimension, linear-fits at different scales were performed in order to capture their main fractal features. Also, we averaged the outcome of $10$ linear fits, distributed over a given interval, in order to improve the precision of the measurements. In all transitions, $D_{\alpha}(p)$ is measured at short length-scales (in particle diameters units) over the interval $r_i\in[1,2]$ with fitting-length equal to $10$, and $r_f\in[11,12]$. At long length-scales, over $r_i\in[10,11]$ with fitting-length equal to $40$, and $r_f\in[50,51]$. For $D_{\beta}(p)$, measurements at medium scales (in particle number) where performed over the interval $r_i\in[10^2,10^3]$ with fitting-length equal to $10^4$, and $r_f\in[1.01\times 10^4,1.1\times 10^4]$. Finally, at large scales, over the interval $r_i\in[10^3,10^4]$ with fitting-length equal to $0.9\times10^5$, and $r_f\in[9.1\times10^4,10^5]$. All fittings to the $D_{\alpha}(p)$ and $D_{\beta}(p)$ data were performed using the \textsl{gnuplot} software.

\vspace{7mm}

{\bf Extended Data} is available in the online version of the paper.\\

{\bf Acknowledgments} The authors gratefully acknowledge the computing time granted on the super\-com\-pu\-ters MIZTLI (DGTIC-UNAM) and THUBAT-KAAL (CNS-IPICyT), and on \hbox{XIUHCOATL} (\hbox{CINVESTAV}) through M.A. Rodriguez (ININ, Mexico). We also acknowledge partial financial support from CONACyT and from VIEP-BUAP, Mexico.\\

{\bf Author Contributions} J.R.N.C.\ carried out all the calculations, performed the analysis of the data and prepared all the figures. V.D.\ supervised the development of the calculations. J.L.C.E.\ supervised the research. All of the authors contributed in the discussion of the results and the preparation of the manuscript.\\

{\bf Author Information} 

Reprints and permissions information is available at www.nature.com/reprints

The author declare no competing financial interests.

Correspondence and requests for materials should be addressed to carrillo@ifuap.buap.mx

\section*{Extended Data}

\begin{figure}[htb]
\renewcommand\figurename{Extended Data Figure}
\setcounter{figure}{0}
\includegraphics[width=\textwidth]{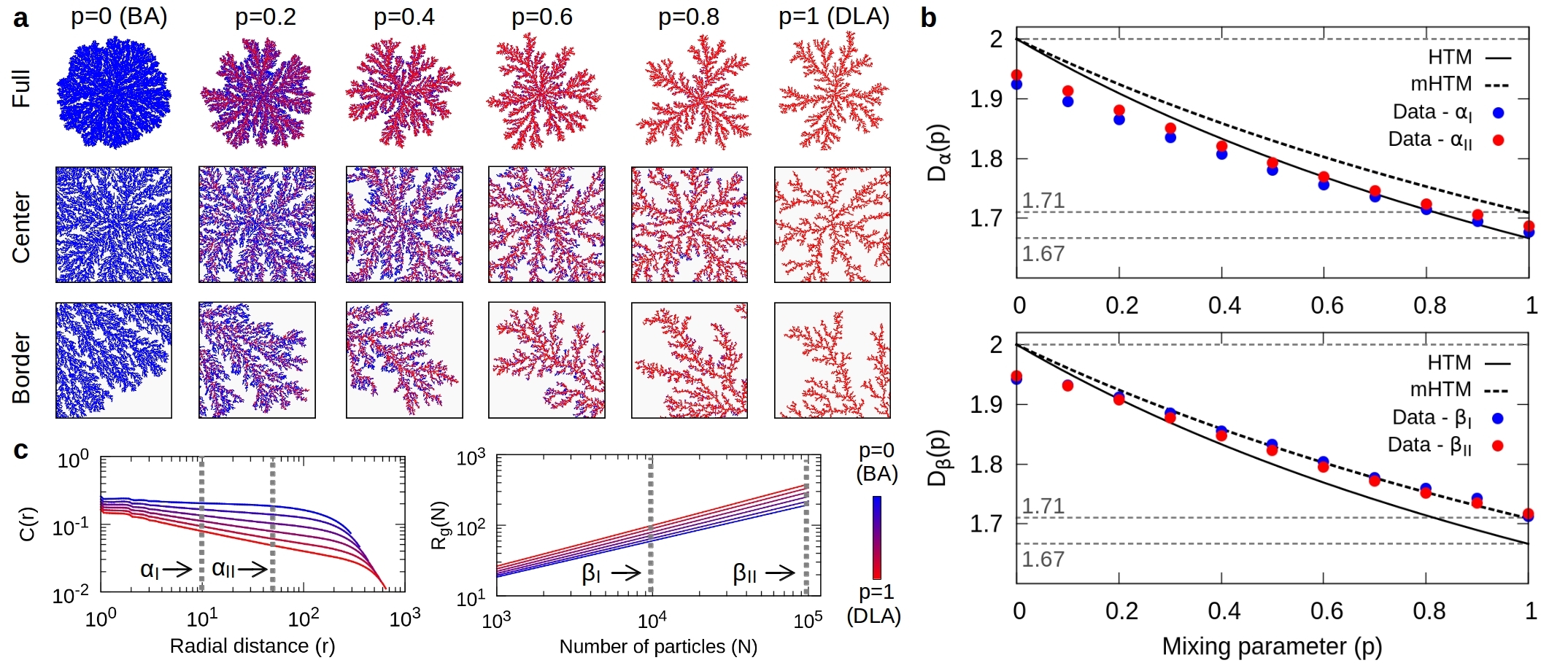}
\caption{\label{EDF1} \textbf{BA-DLA transition.} As explained before, our results suggest that the true fractal dimension of DLA clusters is $D=1.67$, while $D=1.71$ is its first-order approximation. To further support this argument let us analyze the screening-driven BA-DLA transition, characterized by a gradual morphological evolution (\textbf{a}) from stable branching in the BA (blue particles) regime to a unstable tip-splitting in DLA (red particles). Here, the variable $p=N_\mathrm{DLA}/N$ is the ratio of the particles in the cluster aggregated under DLA dynamics and the total number of particles. As $p$ varies from $p=0$ (BA) to $p=1$ (DLA), it controls the morphology of the system. (\textbf{b}) This transition is described by the GHTM equation for $\eta=1$ that, rewritten in terms of $p$, yields $D(d,p,\eta)=(d^2+\eta p)/(d+\eta p)$ with $p=d_w-1$. This expression can be recovered from our equation GD$_1$ for $\chi=1$, $\Lambda=\eta$ and $D_0=d$. As well, the HTM equation is recovered for $\Lambda=1$. In the DLA limit ($p \to 1$), the measured $D_\beta=1.71$ tends to overestimate the value $D=1.67$ given by the HTM equation (solid black curve). This suggests a correction to $\Lambda$ in GD$_1$ can be obtained by setting $D^{(1)}=1.71$ for $p=1$, yielding $\Lambda=0.817$, which is in good agreement (dashed black curve mHTM) with all the data for $D_\beta(p)$. In the DLA limit our equation GD yields $D \approx 1.67$. (\textbf{c}) The self-similarity of the clusters along the transition can be appreciated in the plots of $C(r)$ and $R_g(N)$, displaying well-defined linear behaviours over their respective scales and for different values of $p$. The measured $D_\alpha(p)$ and $D_\beta(p)$ were performed over the indicated regions, respectively.}
\end{figure}

\end{document}